%% file: main_arxiv.tex
\documentclass[twocolumn]{aastex631}
\usepackage{xurl}
\usepackage{amsmath}

\newcommand{\Prot}{P_\mathrm{rot}}
\newcommand{\logg}{\log g}
\newcommand{\sol}{\odot}

\newcommand{\Kepler}{\textit{Kepler}}
\newcommand{\Gaia}{\textit{Gaia}}

\newcommand{\Nhqlcs}{449,913}

\newcommand{\Nnoise}{27,709}

\newcommand{\Nperiods}{32,159} 
\newcommand{\Nkprime}{20,332} 
\newcommand{\Nbkg}{11,219} 
\newcommand{\Nallnew}{9,811} 
\newcommand{\Nkprimeold}{19,650} 

\newcommand{\Nbgfg}{3,224}
\newcommand{\Nbin}{261}
\newcommand{\Npulsators}{608}

\accepted{2025 May 6 to ApJ}

\begin{document}

\title{New Rotation Periods from the Kepler Bonus Background Light Curves}

\correspondingauthor{Zachary R. Claytor}
\email{zclaytor@stsci.edu}

\author[0000-0002-9879-3904]{Zachary R. Claytor}
\affiliation{Space Telescope Science Institute, 3700 San Martin Drive, Baltimore, MD 21218, USA}
\affiliation{Department of Astronomy, University of Florida, 211 Bryant Space Science Center, Gainesville, FL 32611, USA}

\author[0000-0002-4818-7885]{Jamie Tayar}
\affiliation{Department of Astronomy, University of Florida, 211 Bryant Space Science Center, Gainesville, FL 32611, USA}

\begin{abstract}

The \Kepler\ field hosts the best studied sample of field star rotation periods. However, due to \Kepler's large 4\arcsec\ pixels, many of its light curves are at high risk of contamination from background sources. The new \Kepler\ Bonus Background light curves are de-blended using a PSF algorithm, providing light curves of over 400,000 new background sources in addition to over 200,000 re-analyzed \Kepler\ prime targets. These light curves provide the opportunity to search for new rotation periods. Here we apply a convolutional neural network trained on synthetic spot-modulated light curves to regress rotation periods from the \Kepler\ Bonus light curves. We obtained periods for \Nperiods\ total sources, \Nkprimeold\ of which had previously been measured and \Nallnew\ of which are new periods for both \Kepler\ prime and background sources. Our method also detected \Npulsators\ pulsation frequencies from asteroseismic oscillations in red giants. We validate our \Kepler\ prime periods against literature values and present the full period sample. We find excellent agreement with previously-known literature periods, validating deep learning as a viable class of period determination methods. Comparing the periods and light curves of foreground-background pairs, we find that as many as 63\% of periodic background light curves are still blended with the foreground, highlighting limitations of the de-blending technique.

\end{abstract}

\section{Introduction} \label{sec:intro}
Rotation is both a fundamental property of stars and one that is relatively inexpensive to infer in the era of space-based photometric surveys. The \Kepler\ mission \citep{Borucki2010} revolutionized the field of stellar rotation, yielding rotation periods for tens of thousands of stars \citep{McQuillan2014, Santos2019, Santos2021, Reinhold2023}. These rotation periods have empowered studies of stellar ages \citep{Angus2015, Lu2021}, angular momentum evolution \citep[e.g.,][]{vanSaders2016, vanSaders2019, Amard2020, David2022, Avallone2022, See2024}, stellar activity \citep{Mathur2023}, stellar populations \citep{Davenport2017}, Galactic archaeology \citep{Claytor2020}, and exoplanet demographics \citep{Garcia2023}.

Despite the advances enabled by this rich data set, the field of rotation still has many open questions. \citet{McQuillan2013, McQuillan2014} discovered a gap in cool star rotation periods between 10 and 20 days that has since been confirmed in other field star samples (K2: \citealt{Reinhold2020}, ZTF: \citealt{Lu2022}, TESS: \citealt{Claytor2024}). 
\citet{Curtis2019, Curtis2020} found that open cluster stars undergo a period of stalled spin-down between 1 and 3 Gyr. These phenomena may be explained by sequential coupling and decoupling of the radiative core and convective envelope \citep{Lanzafame2019, Spada2020}, but simple models struggle to fully capture the behavior \citep{Curtis2020}. Most investigations would benefit from larger period samples. In particular, the \Kepler\ mission was designed to find Earth-like planets around Sun-like stars, so it has a complicated and biased selection function \citep{Wolniewicz2021}. An unbiased sample would therefore benefit rotation studies further. Attempts have been made using K2 \citep[e.g.,][]{Reinhold2020, Gordon2021} and TESS \citep[e.g.,][]{CantoMartins2020, Avallone2022, Holcomb2022, Kounkel2022, Fetherolf2023, Claytor2024}, but the data quality is not as high as \Kepler\ because of shorter time baselines and uncorrected systematics. {Moreover, TESS is inherently less photometrically precise than the \Kepler\ telescope, limiting rotation detections to higher-amplitude modulations from more active stars.} Answering the open questions about the rotational evolution of main sequence stars will require more periods of the quality of \Kepler\ without the underlying selection bias.

Recently, \citet{Martinez-Palomera2023} reanalyzed the \Kepler\ pixel data using the Linearized Field De-blending method of \citet{Hedges2021} as implemented in the Python package \texttt{psfmachine} \citep{psfmachine}. Whereas previously there were about 200,000 light curves from \Kepler's primary mission, the new KBonus-Background light curves, hereafter ``Kbonus," offer photometric time series of 400,000 new background sources in addition to 200,000 re-analyzed, de-blended \Kepler\ prime targets. These light curves offer an opportunity to potentially triple the number of rotation periods from the well-studied \Kepler\ field. Furthermore, the background sources are not subject to \Kepler's selection function. We therefore have the chance for a large, unbiased, \Kepler-quality period sample, enabling more of the paradigm-shifting science that the mission instigated.

In addition to enhancing the yield of periods from \Kepler, the de-blended light curves offer a testbed to assess the susceptibility of period-finding methods to background contamination. This is even more important to the mission of the \textit{Transiting Exoplanet Survey Satellite} \citep[TESS, ][]{Ricker2015}, which features larger 21\arcsec\ pixels. Launched in 2018, TESS is still relatively new, and rotation studies are just beginning to hit their stride \citep{CantoMartins2020, Avallone2022, Holcomb2022, Kounkel2022, Fetherolf2023, Claytor2024}. Because it is an all-sky survey, TESS provides millions of light curves to be analyzed for rotation signals, and already over 100,000 short periods ($< 12$ days; {aggregated from the aforementioned studies}) have been obtained. However, systematics related to the satellite's orbit have slowed the recovery of longer periods using conventional methods. {Efforts have been made to circumnavigate TESS's complications using systematics-resistant periodograms \citep{Hedges2021}, more robust pixel corrections \citep{Hattori2022}, and better light curve stitching routines \citep{Palakkatharappil2024, Garcia2024}, but even a one-size-fits-most solution remains evasive.} \citet{Claytor2024} used a convolutional neural network (CNN) with simulated training sets to regress periods up to 80 days in TESS's southern continuous viewing zone (CVZ). While this was a crucial step to increase the number of available rotation periods, the lack of overlap between the TESS CVZs and previous studies results in an absence of adequate real-world validation data for this deep learning approach. The \Kepler\ mission has a 10-year advantage over TESS, making Kbonus the perfect validation set for our deep learning methods.

In this paper we apply the deep learning framework of \citet{Claytor2024} to the Kbonus light curves in order to estimate rotation periods, providing the first look at rotation in the background sources and additional validation of the method. We simulate 1 million new training light curves with the cadence and baseline of the \Kepler\ prime mission and combine them with a selection of flattened Kbonus light curves to emulate noise and systematics. We train and evaluate the same CNNs on the synthetic training set and then evaluate the best-performing model on the full Kbonus data. Highlighting structure in the predicted parameters, we identify a new way to filter out bad predictions. {After examining the distribution of rotation periods, we compare our estimates to the literature. We use the periods measured from foreground-background pairs in addition to light curve cross-correlation analysis to assess the quality of de-blending. Finally, we identify as source confusion candidates cases where previous period measurements were likely to be from the background star. We present our rotation period estimates, foreground-background pair identification, and cross-correlation statistics in this first step toward an unbiased analysis of stars in the \Kepler\ field.}

\section{Data} \label{sec:methods}
We used the Kbonus light curves for this analysis. The light curves and corresponding source catalogs are available from the Mikulski Archive for Space Telescopes (MAST) as a High Level Science Product\footnote{DOI: \dataset[10.17909/7jbr-w430
]{http://doi.org/10.17909/7jbr-w430
}} \citep[HLSP,][]{kbonus}. We mirrored the full 4.8 TB data set onto the HiPerGator supercomputer at the University of Florida\footnote{{UFIT Research Computing,
University of Florida: \url{https://www.rc.ufl.edu}.}}

{The Kbonus light curves were created by modeling aberrations and using those time-dependent models to perturb or correct the instrument point-spread function (PSF) for each source, using \Gaia\ Data Release 3 \citep[DR3,][]{Gaia2023} sources as input. Fluxes were computed using the mean PSF as well as the time-dependent corrected PSF \citep[for more information on the PSF models and the perturbations thereof, see, respectively, Sections 2.4 and 2.5 of][]{Martinez-Palomera2023}}. 
{In addition to corrected PSF fluxes, the light curves also feature flattened PSF fluxes specialized for identifying exoplanet transits (for these, a 2 day window B-spline was fit to the data and removed).}
We used a subset of flattened PSF fluxes as noise templates for our training data (discussed in Section~\ref{sec:training}), and we used the corrected PSF light curves for our rotation search. In both cases, we began with the stitched light curve, removed cadences with any quality flags (using \texttt{Lightkurve}, \texttt{quality\_bitmask=`hardest'}), removed 3$\sigma$ flux outliers, and kept all light curves with at least 1 quarter of PSF photometry. Seven stars\footnote{\Kepler\ Input Catalog (KIC) IDs 2437601, 5111608, 5111668, 7529265, 10404875, 10404896, and \Gaia\ DR3 2076488530099163776.} had only one cadence left using the ``hardest" quality mask; we relaxed the bitmask to ``default" for these. This left us with \Nhqlcs\ high-quality light curves. The Kbonus source catalog includes other useful quality-control metrics, which we used to spot-check our rotation analysis rather than for selection criteria. These include:

\begin{itemize}
    \item PSFFRAC: the estimated fraction of the PSF captured in the target pixel file.
    \item PERTRATI: the ratio of the mean PSF flux to the perturbed model PSF flux.
    \item PERTSTD: the ratio of the PSF flux standard deviation to the perturbed model PSF flux standard deviation.
\end{itemize}

\begin{figure*}
    \centering
    \includegraphics{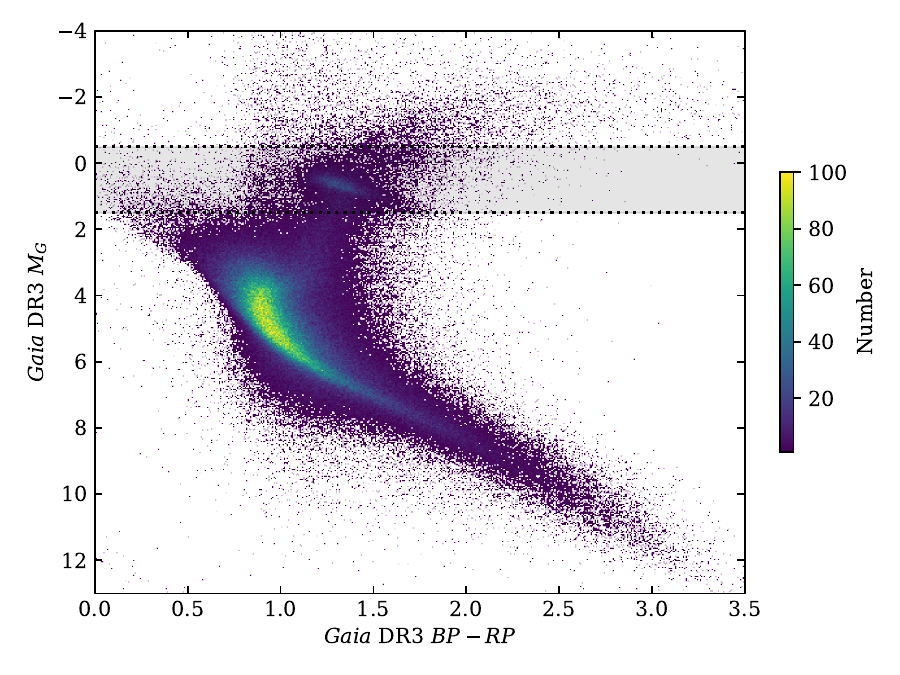}
    \caption{The \Gaia\ color-magnitude diagram of 425,191 Kbonus targets with a high-quality PSF light curve spanning at least 1 quarter. The horizontal gray band highlights the magnitude range (roughly centered on the red clump) we used to select light curves to be combined with simulated light curves for our training set. Missing from this plot are 23,636 stars missing \Gaia\ $G$, $BP$, $RP$, or parallax measurements and 1,086 stars outside the plotting range.}
    \label{fig:cmd}
\end{figure*}

We also queried the \Gaia\ DR3 source catalog \citep{Gaia2023} using \texttt{Astroquery} to obtain parallax measurements for our targets. We then computed the absolute \Gaia\ magnitude $M_G$:
\begin{equation} \nonumber
    M_G = g + 5 \log p - 10,
\end{equation}
where the parallax $p$ is in milliarcseconds.  Figure~\ref{fig:cmd} shows the \Gaia\ color-magnitude diagram (CMD) of our high-quality targets, with the stars we used as noise examples in the training set (described in Section~\ref{sec:training}) highlighted by the horizontal gray band. In this work, we use the \Gaia\ photometry only to select stars to be noise examples for the training set, and to remove pulsating giants from our rotation period sample. Neither of these selections is significantly affected by reddening or extinction, so we ignore these effects.


\section{Training Data} \label{sec:training}

To build our training set, we used simulated rotational light curves combined with real photometric noise from the Kbonus light curves. Following \citet{Claytor2022, Claytor2024}, we used the new version 1.0 of \texttt{butterpy} \citep{butterpy1.0} to simulate 1 million rotational light curves with \Kepler's 4-year baseline sampled at 30-minute cadence. Version 1.0 introduced an object-oriented interface, computation speedups, new and improved documentation, and a more accurate Solar normalization for star spot emergence.

Our simulation input parameters were sampled from the same distributions as \citet{Claytor2022}, uniformly sampling rotation periods from 0.1 to 180 days and activity levels log-uniformly from 0.1 to 10 times Solar. We partitioned the data with an 80\%/10\%/10\% split for training, validation, and test sets.

A representative training set must include noise and systematics similar to the real light curves. We therefore used a subset of flattened PSF light curves to encapsulate the noise properties of the Kbonus data. Targeting red clump stars, which generally rotate slowly \citep[e.g.,][]{Tayar2018, Daher2022, Patton2023} and should have little-to-no detectable rotation signatures (\citealt{Ceillier2017} detected rotation in 2\% of red giants and 15\% of $M < 1.1 M_\sol$ red clump stars), we selected stars with absolute \Gaia\ magnitude $-0.5 < M_G < 1.5$ and defined $BP$ and $RP$ magnitudes, which yielded \Nnoise\ stars at or near the red clump with high-quality PSF light curves. 

We note that the use of flattened PSF photometry means that our training light curves lack any trends longer than 2 days which the corrected PSF light curves retain. This would pose a critical flaw in the procedure for a mission like TESS, whose light curves are often dominated by scattered light systematics on stellar rotation timescales \cite[e.g.,][]{TESSHandbook}. However, due to \Kepler's Earth-trailing orbit and 90-day quarter length, the light curves are far less affected by such systematics, which typically occur on timescales longer than those of rotation \citep[e.g.,][]{Santos2019}. 

\begin{figure}
    \centering
    \includegraphics[width=\linewidth]{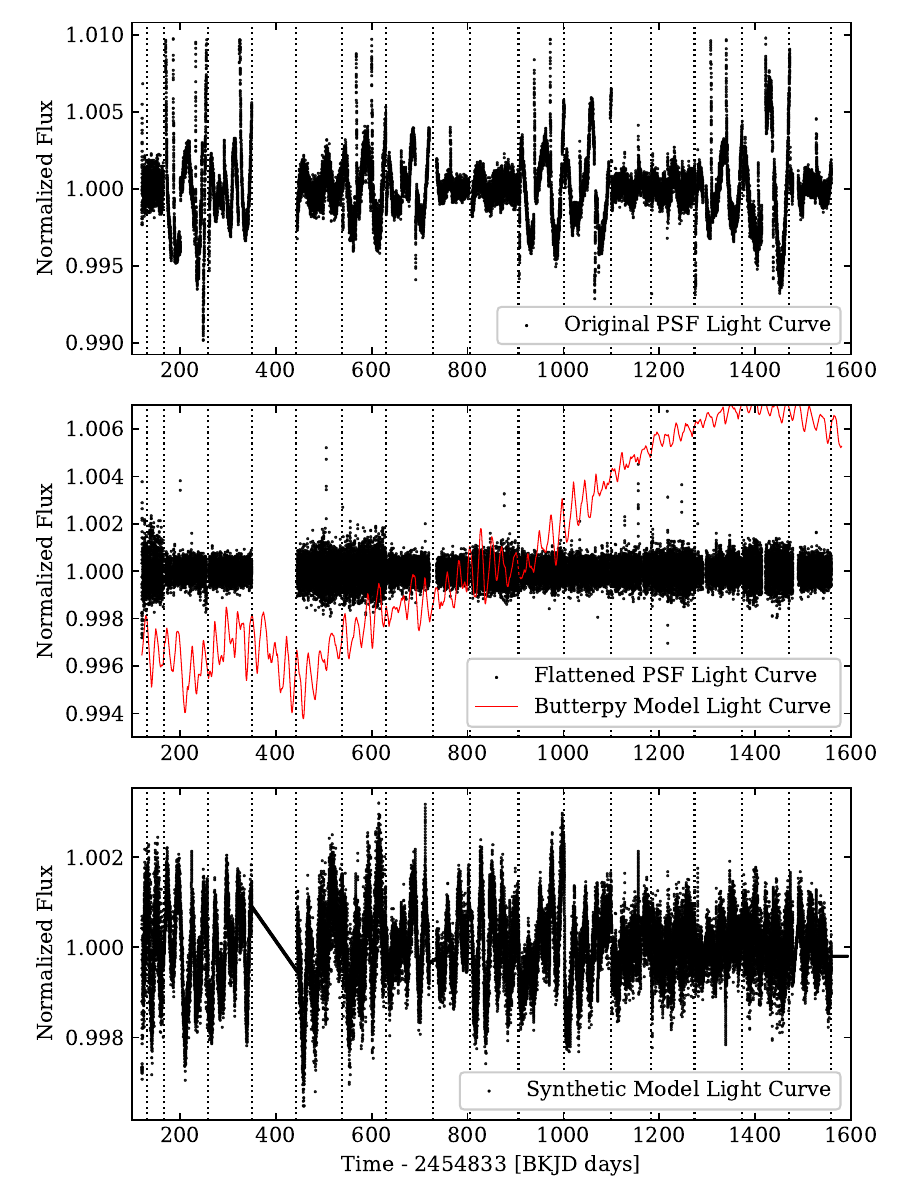}
    \caption{Example light curves of KIC 2016676, a Kbonus target used as a noise example, and simulation 138945 from our training set. The vertical dotted lines in all panels correspond to \Kepler\ quarter divisions. \textit{Top}: the Kbonus PSF light curve with flagged cadences removed, 3$\sigma$ outliers rejected, and flux median-normalized. \textit{Middle}: the flattened PSF light curve of the star (black) and the \texttt{butterpy} model light curve (red) to be combined. The model has a rotation period of 25 d. \textit{Bottom}: the combined, synthetic training light curve, which has been divided into \Kepler\ quarters, re-normalized, and stitched, hence the lack of the long-term trend seen in the raw model curve.}
    \label{fig:lcs}
\end{figure}

We shuffled and likewise partitioned the noise light curves 80\%/10\%/10\% to be combined with our training, validation, and test light curves. This way, each noise example would be used roughly 37 times with no cross-contamination between the three training partitions. To combine the noise and simulated light curves, we followed the following procedure:
\begin{enumerate}
    \item Split both the simulated and noise light curves into 90-day segments.
    \item Median-normalize each segment.
    \item Interpolate each simulated segment to the cadences of its corresponding noise segment.
    \item Multiply the interpolated simulation segment by the noise segment.
    \item Stitch the product segments together.
    \item Linearly interpolate to fill gaps in time.
\end{enumerate}
Multiplying the light curve segments rather than adding them preserves the ratio of rotational amplitude to noise amplitude and ensures a realistic signal-to-noise ratio. Figure~\ref{fig:lcs} illustrates an example light curve at various steps of the noise combining pipeline. 

Finally, we used a binned Morlet wavelet transform \citep{Torrence1998} of the combined light curves as the input representation to our CNN. We transformed the light curves using the continuous wavelet transform implemented in \texttt{SciPy} \citep{Scipy2020} with the power spectral density correction of \citet{Liu2007}, plotted the power spectrum for $0.1~\mathrm{d} \leq \Prot \leq 180~\mathrm{d}$ for the entire 4-year baseline, binned the spectrum to a 64$\times$64 array, and min-max scaled the array to integers in the range [0, 255] for memory efficiency.

{The full set of simulated light curves, their corresponding wavelet transforms, and the input parameters used to create them with \texttt{butterpy} are available from MAST under the HLSP ``Stellar Magnetism, Activity, and Rotation with Time Series" (SMARTS)\footnote{DOI: \dataset[10.17909/davg-m919
]{http://doi.org/10.17909/davg-m919
}}}.
\section{CNN}
\label{sec:cnn}
CNNs exhibit local connectivity, which makes them efficient at learning features from data that are spatially correlated \citep{Lecun2015}. This makes them ideal tools in many astrophysical contexts, from one-dimensional sequences like spectra \citep{Sharma2020}, light curves \citep{Pearson2018, Zucker2018}, to two-dimensional structures like images \citep{Hezaveh2017}, multivariate distributions and power spectra \citep{Claytor2022, Claytor2024}, and even images of one-dimensional data \citep{Hon2018, Hon2021}. We employ a CNN to take advantage of the high spatial correlations in light curves and their power spectra.

We used the same CNN architecture as \citet{Claytor2024}, {which consists of three convolution blocks followed by three fully connected (FC) layers. Each convolution block comprises a 2D convolution layer, 1D max pooling, rectified linear (ReLU) activation, and 10\% dropout. The convolution layers have increasing numbers of trainable kernels which serve as feature extractors. The output of the convolution block is a series of feature maps, which are flattened and passed to the FC layers, two of which have ReLU activation and 10\% dropout. The final FC layer has softplus activation and produces two outputs: the rotation period and uncertainty. For a more detailed overview, see Table 3 of \citet{Claytor2024}.}

For training, we used the Adam optimizer \citep{Kingma2014} with negative log-Laplacian loss:
\begin{equation} \label{eq:loss}
    \mathcal{L} = \ln\left(2b\right) + \frac{|P_\mathrm{true} - P_\mathrm{pred}|}{b},
\end{equation}
where $b$, the median absolute deviation, is taken to represent the uncertainty. This loss function enables the estimation of uncertainty along with the rotation period \citep{Claytor2022}. 
We fit the neural network to the training data with a batch size of 100 {wavelet transforms}, iterating until the validation loss reached a local minimum or plateau. Tracking the validation loss this way allowed us to cease training before the CNN overfit the training data. 

As in \citet{Claytor2024}, four architectures with increasing numbers of convolution filters were trained. Assessing performance by median relative error and by overall accuracy, all four architectures performed roughly equally on the held-out test partition. However, the simplest {(the ``A" configuration of \citealt{Claytor2024}, which had three convolution layers with 8, 16, then 32 filters)} overfit the least and had the least structure in the predicted period distribution. We used this model to estimate rotation periods from the Kbonus light curves.

\begin{figure}
    \centering
    \includegraphics[width=0.9\linewidth]{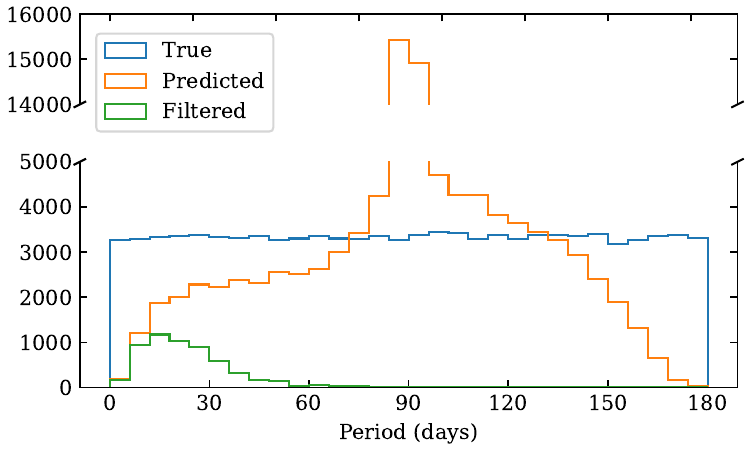}
    \includegraphics[width=0.9\linewidth]{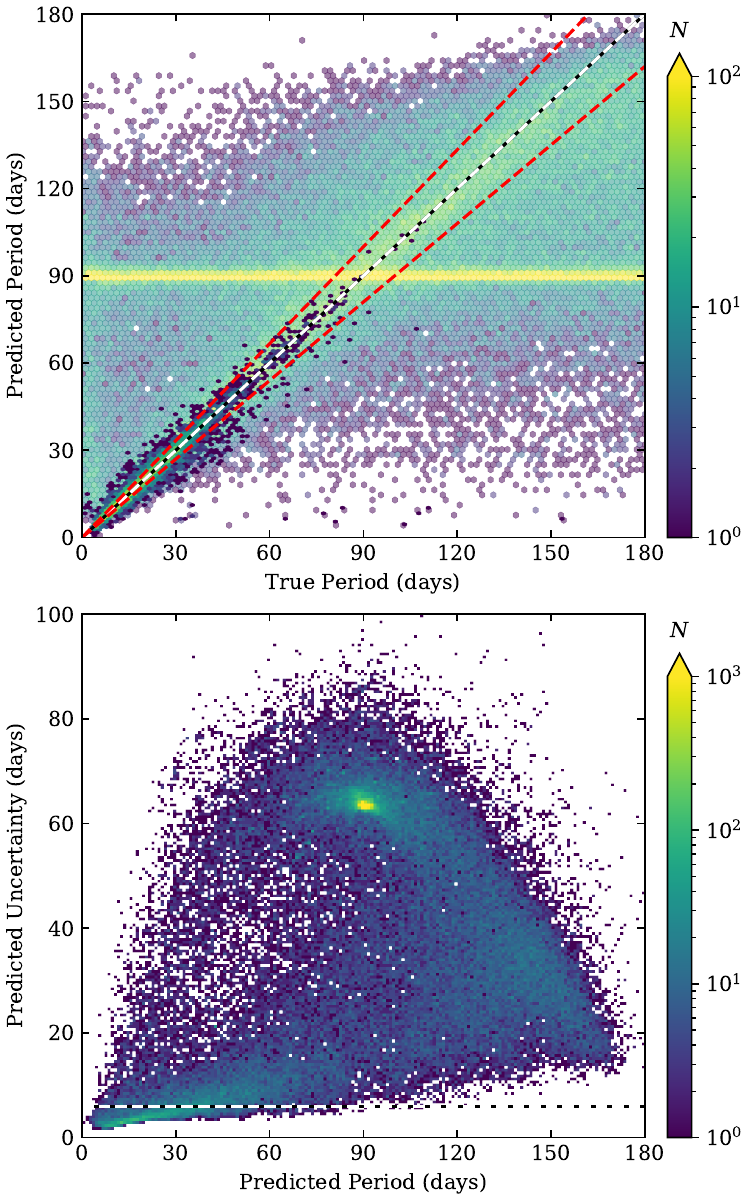}
    \caption{Distributions of CNN-predicted parameters and comparison to the true, underlying values for the simulated test data, {with an uncertainty-based quality cut ($\sigma < 6$ d) illustrated}. \textit{Top}: histograms of the true, predicted {and filtered} periods. The true periods are distributed uniformly between 0.1 and 180 days, but the predicted periods have a sharp peak at the distribution median of 90 days. Note the broken y-axis. {The strict quality cut filters out most of the data, but removes the uncertain 90 d predictions.} \textit{Middle}: predicted vs. true period {for the full sample (low opacity) and filtered sample (high opacity)}. 29\% of test examples are recovered with 10\% or greater accuracy (red dashed lines). {After filtering, 66\% (87\%) are recovered within 10\% (20\%) accuracy}. \textit{Bottom}: predicted uncertainty vs. period highlighting the sharp peak at $P_\mathrm{pred} = 90$ d, which is highly uncertain. {The strict $\sigma < 6$ d cut is illustrated by the horizontal dashed line.}}
    \label{fig:predictions}
\end{figure}

Figure~\ref{fig:predictions} illustrates the CNN performance on the held-out test sample. While the true periods are distributed uniformly (top), the predictions are underrepresented at the edges of the distribution, but largely overrepresented at 90 days. Objects spanning the full range of true periods are erroneously predicted to have 90 day periods (middle), but these objects are also highly uncertain (bottom), making them easy to remove using a quality cut. With no cuts, 29\% of test periods are recovered with 10\% error or less, and 47\% of periods are recovered to within 20\% accuracy. This is nearly identical performance to using the same framework with TESS light curves: \citet{Claytor2022} recovered 28\% and 45\% of TESS-like test targets to within 10\% and 20\% accuracy, respectively. Identical performance between two data sets with different quality could mean that the limiting factor is the CNN model and not the data, suggesting that more advanced architectures may be able to extract more or better rotation periods from the data.  {Such architectures might employ additional convolution layers or layers with more convolution kernels, features like batch normalization for more efficient training, or additional input features such as light curve or periodogram statistics.} We leave the testing of more complex architectures to future work.

{Based on the CNN performance on the real Kbonus data (further discussed in the next section), we determined that restricting analysis to the predictions with uncertainty less than 6 days provided the highest quality sample with the least data loss. We have included illustrations of that cut in Figure~\ref{fig:predictions}. While the cut is very strict, removing nearly 95\% of the simulated test data, the periods that remain are highly accurate, with 87\% of the periods recovered to 20\% accuracy or better.}

As in \citet{Claytor2022, Claytor2024}, the CNN performance is worse toward the edges {of the period range. This compression of the prediction range is inherent to machine learning regression under measurement uncertainty \citep{Ting2024}.} As a result, we struggle to recover rotation periods less than about 5 days. While we could circumvent this using different training sets with smaller period ranges \citep[as in][]{Claytor2024}, there are a variety of other methods optimized for the estimation of short periods \citep{Luger2021, Holcomb2022, Reinhold2022}. {Future work should use a combination of methods---such as more conventional Fourier analysis for short periods, and deep learning for long periods---to achieve a more comprehensive rotation analysis.}

\begin{figure}
    \centering
    \includegraphics[width=\linewidth]{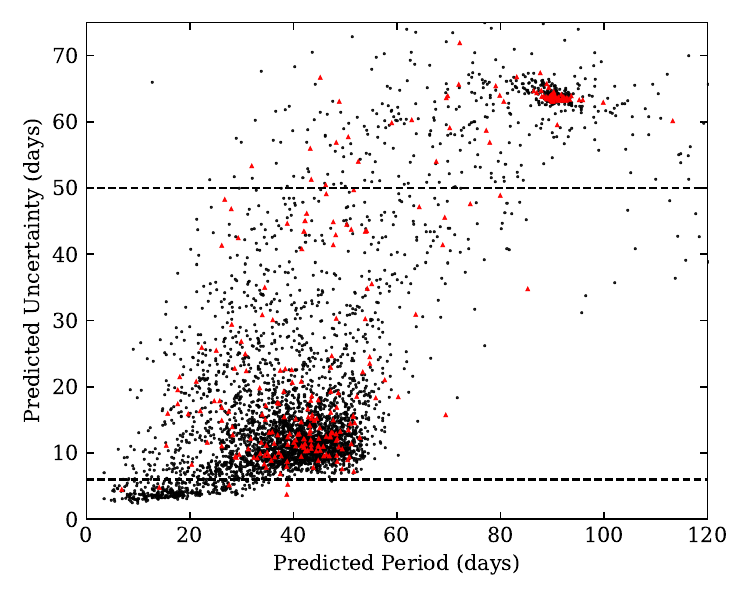}
    \caption{Predicted uncertainty vs. period for a representative 1\% of our full sample (black circles) and of targets that were used as noise examples in the training set (red triangles). The predictions fall roughly into three clusters, which we have separated by black dashed lines at uncertainties of 6 and 50 days. The bottom sample represents the cleanest, highest-fidelity set of periods, while the middle has high contamination of incorrectly-predicted periods. The top cluster represents targets for which the CNN could not identify a period signal and assigned the median value of 90 days with high uncertainty.}
    \label{fig:real_preds}
\end{figure}

\pagebreak

\section{Results \& Discussion} \label{sec:results}

\subsection{Predicted Periods and Uncertainties}

\input{periods_table}

We evaluated the simplest CNN model on the wavelet transform images {of Kbonus light curves}. We present the periods, their associated uncertainties, flags, and relevant Kbonus and \Gaia\ source parameters in Table~\ref{tab:periods}. Figure~\ref{fig:real_preds} shows a representative sample of predicted periods and uncertainties. Black points represent our full Kbonus sample, while red triangles represent targets that we used as noise examples for training. The predictions roughly cluster in three uncertainty regimes, separated by horizontal dashed lines at uncertainties of 6 and 50 days. 

We evaluated the more complex CNN models on the wavelet transform images as well. These models returned more periods that we would consider ``good" with small absolute and fractional predicted uncertainties. However, the more complex models also had higher contamination of clearly bad periods with small uncertainties, with more scatter between the three uncertainty regimes highlighted in Figure~\ref{fig:real_preds}, making the distinction between good and bad predictions less clear. Prioritizing purity and reliability of the period sample over completeness, we adopted the period predictions from the simplest CNN model.

The highest-uncertainty cluster of Figure~\ref{fig:real_preds} consists almost entirely of stars with fewer than three quarters of observations. As with the test set shown at the bottom of Figure~\ref{fig:predictions}, the CNN could not identify rotation periods for these targets and assigned most of them the median value of 90 days with high uncertainty. The bottom two clusters {in Figure~\ref{fig:real_preds} are not seen in the test set recovery}, which has no clear bimodality. We suspect that the cluster near 40-day period and 10-day uncertainty may be related to systematics around downlink times. Because we used the flattened PSF light curves to construct our training set, these systematics are not present in our {noise model}, and the CNN may mistake the systematics for astrophysical periodicity. We leave investigation into these systematics for future work and restrict ourselves to the \Nperiods\ targets with uncertainties less than 6 days for our remaining analysis.

The stars used as training noise examples (red triangles) are significantly underrepresented in the lowest-uncertainty cluster. These targets make up 10\% of the highest-uncertainty cluster, 6\% of the middle, but only 1\% of the bottom sample. One possible explanation is that, because these targets are from the red clump, little-to-no rotation signal is present in any of the light curves \citep{Ceillier2017}, and the CNN assigns them high uncertainty as expected. Another possibility is that the CNN has overfit the noise from the training set. Seeing the same noise properties with different injected rotation signals may confuse the CNN such that it can make no reliable prediction for these targets if there is indeed rotation signal present in the real, unflattened light curve.
{Perhaps the likeliest reason is that the red clump light curves have worse photometric precision than stars elsewhere on the CMD \citep[see Section 3.1.2, including Figures 8 and 9, of][]{Martinez-Palomera2023}. Why these light curves should be noisier is not immediately clear. It may be that the clump stars appear relatively noisy because their oscillation periods \citep[$\sim$9 hours,][]{Kjeldsen1995} are close to the window length (6 hours) used for the combined differential photometric precision (CDPP) measurement, and the measurement mistakes oscillation for noise. If this is the case, CDPP could be an inexpensive tool to detect oscillating giants in other surveys such as TESS. We leave this exploration to future work.}

\begin{figure*}
    \centering
    \includegraphics[width=0.9\linewidth]{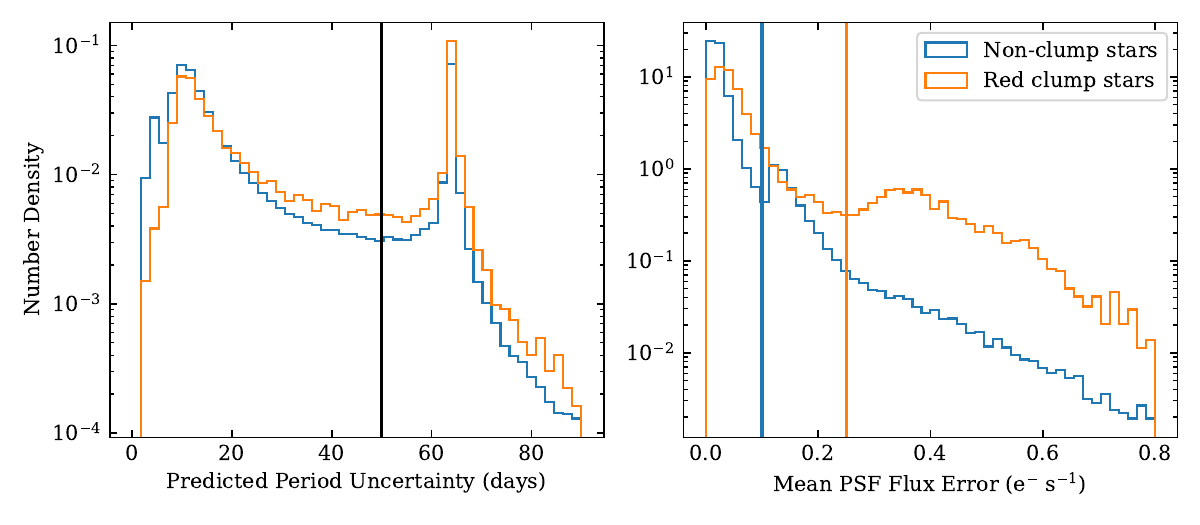}
    \caption{Distributions of predicted period uncertainty (left) and light curve flux error (right) for red clump (orange) and non-clump stars (blue). The vertical lines represent the positions of minima separating the two modes of each distribution. In the left panel, the minima occur at the same uncertainty, hence the single black line. The flux error distributions have different minima, and the lines denoting the minima are colored to match their corresponding histograms. The clump stars have worse photometric precision than non-clump stars, which directly translates to higher uncertainty in their predicted rotation periods.}
    \label{fig:clump}
\end{figure*}

{We expect the worse photometric precision of red clump light curves to result in the inflation of the uncertainties of clump star periods estimated by the CNN. To test this hypothesis, we made histograms of the predicted period uncertainty and mean flux error (a proxy for photometric precision) for the red clump and non-clump stars in the full, unfiltered sample. Figure~\ref{fig:clump} shows the histograms, which emphasize the bimodality of each distribution. We then used the \texttt{find\_peaks} function in \texttt{scipy.signal} to identify the minima between the two modes of each distribution, testing a range of bin sizes to suppress error from the discretization of counts. We used the locations of the minima to divide each distribution into a high- and low-value sample. Finally, we quantified the relative representation of sources in each sample using the fraction of sources in the high-value sample, then comparing that fraction for clump stars to the fraction for non-clump stars. 14\% of clump stars were in the high-noise sample, compared to 8\% of non-clump stars, meaning that red clump stars are over-represented in the high-noise sample by a factor of 1.75 (14\%/8\%). Similarly, 32\% of clump stars were in the high-uncertainty sample, compared to 21\% of non-clump stars, with an over-representation ratio of about 1.52. The worse precision alone is therefore sufficient to account for the worse period recovery among clump stars. To maintain the fidelity of our sample, we remove the red clump stars from further analysis.}

At this point we looked at the PSFFRAC, PERTRATI, and PERTSTD values to determine whether quality cuts needed to be made. Based on the parameter distributions in the Kbonus source catalog, we considered light curves with PSFFRAC $> 0.7$, $0.99 <$ PERTRATI $< 1.01$, and PERTSTD $< 0.05$ to be high-quality {(i.e., at least 70\% of the flux was contained by the PSF model, and perturbing the PSF caused no more than 1\% deviation in mean flux and no more than 5\% change in the flux standard deviation, respectively)}. Of the \Nperiods\ light curves for which we found periods, 22,970 had PSFFRAC $> 0.7$ (19,456 foreground and 3,514 background), all targets had high-quality PERTRATI save for one foreground star, and all but seven (one foreground and six background) PERTSTD $< 0.05$. Combining all cuts, 22,967 (19,455 foreground and 3,512 background) targets had high-quality light curves. Despite the large fraction (29\%) of targets with low-quality PSFFRAC values, applying quality cuts to our data did not significantly affect any of our results. We therefore retain all of our targets but reproduce the quality metrics in Table~\ref{tab:periods} for transparency.

\subsection{The Kbonus Period Distribution}
\label{sec:distribution}

\begin{figure*}
    \centering
    \includegraphics[trim={0.8cm 0.3cm 0.3cm 0.3cm},clip,width=\linewidth]{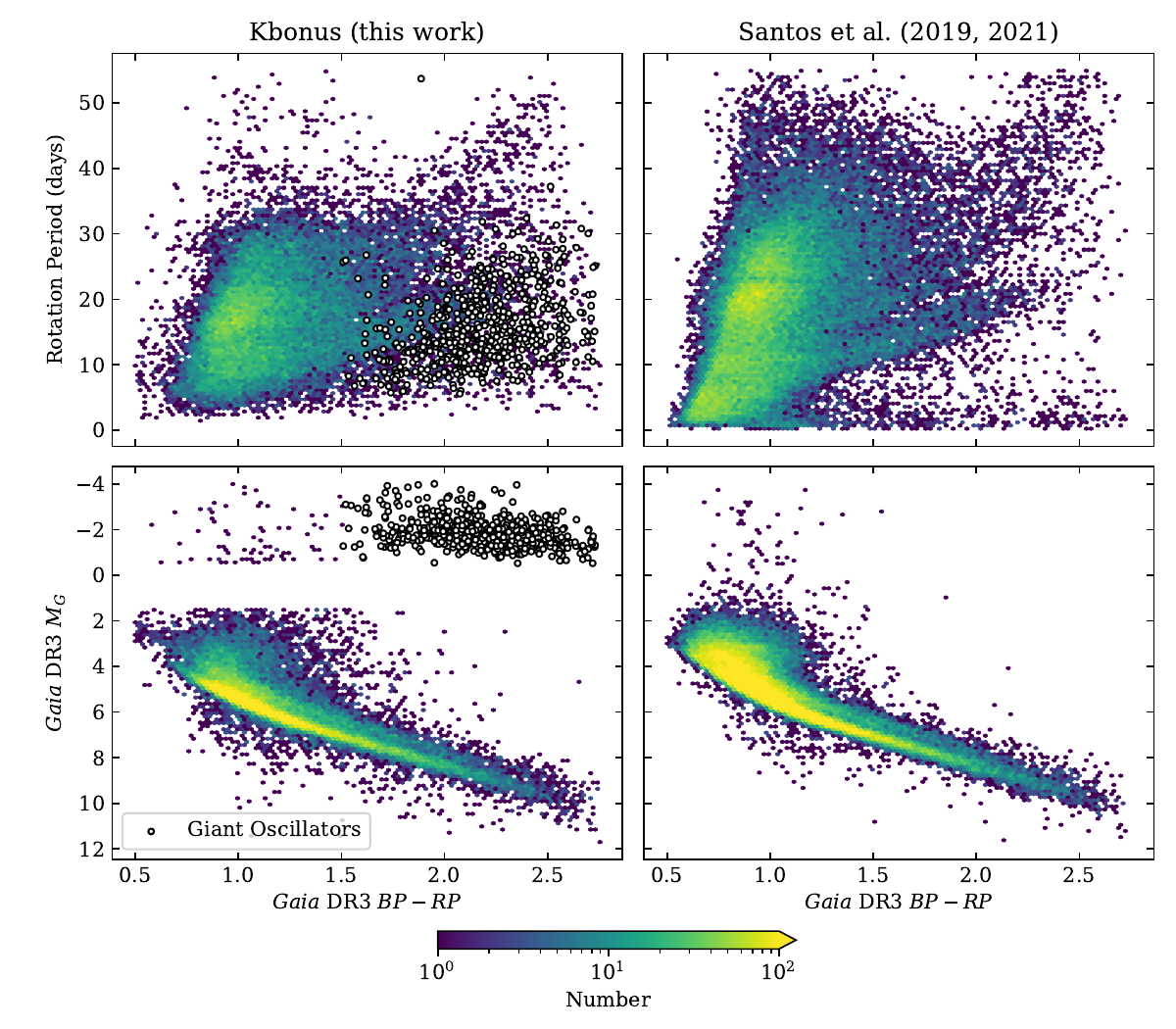}
    \caption{\textit{Left}: The period distribution and \Gaia\ CMD of 30,513 of our targets with high-fidelity periods. The bins are colored by number density and are logarithmically scaled to highlight fine structure. \Npulsators\ oscillating giant {candidates} are shown as white circles. The empty horizontal band in the CMD is centered on the red clump, where our training light curves were taken from, so we exclude them from analysis. Not shown are 569 targets with undefined $RP$ magnitudes and 1,032 targets outside the plotting range. \textit{Right}: The \Kepler\ rotation sample of \citet{Santos2019, Santos2021} shown in the same axis ranges and color scales for comparison. 335 stars, mostly sub-giants and M dwarfs, have periods longer than 55 days and are outside the plotting range.}
    \label{fig:period_dist}
\end{figure*}

After using the predicted uncertainty to select high-fidelity periods, we can view the distribution of Kbonus rotation periods as a function of \Gaia\ color. Figure~\ref{fig:period_dist} shows the period distributions and \Gaia\ CMDs of our Kbonus sample and the \Kepler\ stars of \citet{Santos2019, Santos2021}. The top left panel shows our period distribution, which includes our rotation periods for \Nkprime\ \Kepler\ prime targets and for \Nbkg\ background sources. The period distribution closely resembles that of the \Kepler\ prime distribution, including the gap in rotation periods around 20 days for cool stars as well as the ``tail" of slowly-rotating M dwarfs. The distribution also exhibits a short-period edge where stars converge onto a well-behaved spin-down sequence \citep{Curtis2020}. 

The region below the short-period edge and redder than $BP - RP = 1.5$ {likely contains some rapidly rotating close binary main sequence stars \citep[][also visible in the CMD as a binary main sequence]{Simonian2019} or stellar merger products. However, most of the sources in this region are oscillating giant candidates}. {We identified oscillating giant candidates as stars with $M_G < 0$, $BP-RP > 1.5$, and a period detected by the CNN, since periodicity in these stars is very unlikely to be from rotational modulation. We have flagged all our oscillator candidates in Table~\ref{tab:periods} for community follow-up.} Importantly, the detection of oscillations means that our CNN is not robust against contamination from periodic but non-rotational sources. Conversely, the machine we designed as a rotation detector also happens to detect seismic oscillations (and presumably other forms of periodicity), but quantifying the sensitivity to these other periodic sources is beyond the scope of this work. 

The main differences between our period distribution and that of \citet{Santos2019, Santos2021} are at the very short-period end ($P_\mathrm{rot} < 2$ d) and the somewhat slowly rotating blue end ($1 < BP - RP < 2$, $P_\mathrm{rot} > 30$ d). In particular, our distribution lacks rapidly rotating stars, which are near the edge of the range of the CNN predictions. Near the edges, any uncertainty in the estimate causes predictions to be biased toward the midpoint of the range and away from the edges \citep[for further discussion, see][]{Claytor2020, Claytor2022}. As a result, our CNN struggles to recover periods at the very short end. On the blue end we failed to recover many periods longer than 30 days. In the Santos et al. sample, these stars are mostly old, late G and K dwarfs, along with a few sub-giants. Such stars rotate slowly and tend to have small variability amplitudes that often get lost in the noise, making them more difficult to detect. Additionally, Santos et al. searched for periods using three different light curves for each star, where each light curve was filtered to optimize recovery in different period ranges. This enhanced their sensitivity to rotation signals even in the presence of noise. We performed no such filtering on the Kbonus light curves, inhibiting our detection of slow rotation in low-activity stars. 

\subsection{Comparison to Literature Periods}

\begin{figure*}
    \centering
    \includegraphics[width=\linewidth]{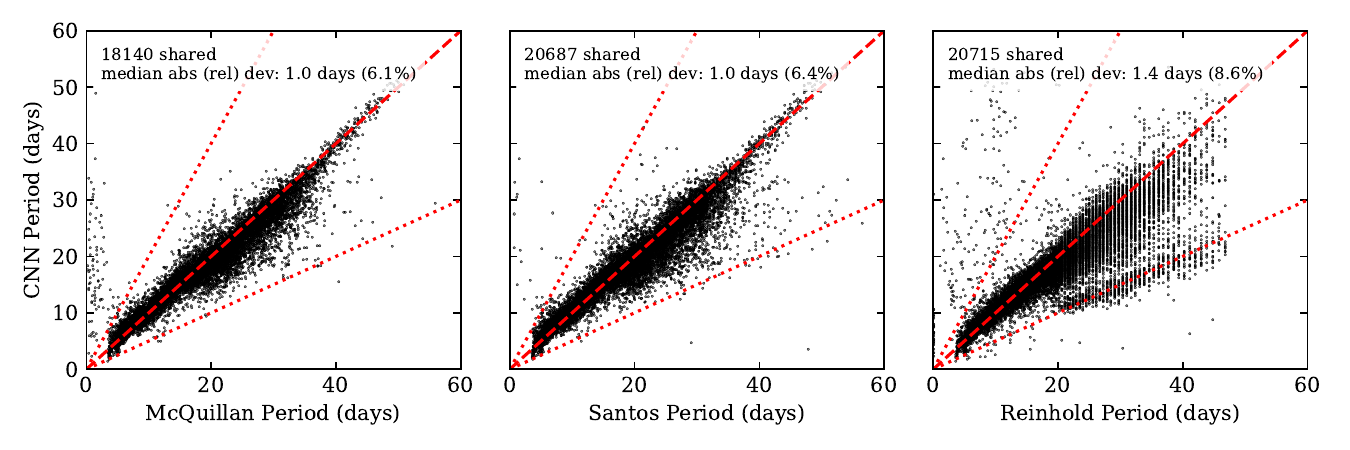}
    \caption{Comparison of our CNN rotation period to the rotation periods from the \Kepler\ prime analyses of \citet[][left]{McQuillan2014}, \citet[][middle]{Santos2019, Santos2021}, and \citet[][right]{Reinhold2023}. Each sample shares roughly 20,000 period measurements with ours, and our periods have a median absolute deviation of $\sim$1 day and relative deviation of $\sim$6--9\% in comparison to the literature. The red lines denote equality (dashed) and factor-of-two aliases (dotted).}
    \label{fig:comparison}
\end{figure*}

Of our \Nperiods\ CNN-estimated periods, \Nkprime\ are of \Kepler\ prime targets. Of these, \Nkprimeold\ overlap with either \citet{McQuillan2014}, \citet{Santos2019, Santos2021}, or \citet{Reinhold2023}. Figure~\ref{fig:comparison} compares our rotation periods of \Kepler\ prime targets with these literature sources. The periods agree very well, with median absolute deviations of roughly 1 day and median relative deviations of 6--9\%.

Interestingly, our periods do not show half- or double-period aliases with the autocorrelation periods of \citet{McQuillan2014} or the combined autocorrelation and wavelet periods of \citet{Santos2019, Santos2021}. However, for a small number of cases we measured half the period measured by \citet{Reinhold2023}, who used the gradient of the power spectrum \citep{Reinhold2022}. As \citet{Reinhold2023} discuss, their method occasionally recovers twice the true period, especially when the light curve modulation is highly periodic. We interpret the lack of aliasing as strong verification of our CNN recovering the correct period.

\begin{figure*}
    \centering
    \includegraphics[width=0.8\linewidth]{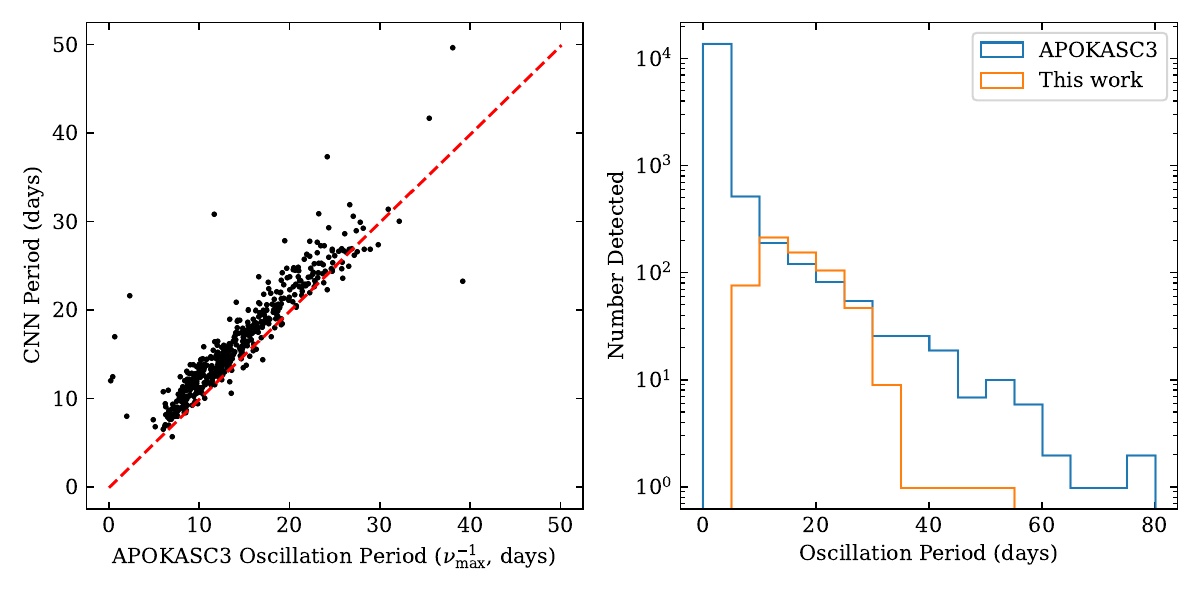}
    \caption{\textit{Left}: Our CNN-inferred period versus the APOKASC3 frequency of maximum power \citep{Pinsonneault2024} for 480 stars in common between the two samples. Our periods exhibit a 2-day bias compared to those from APOKASC3, but otherwise show good agreement. This offset is likely due to the granulation background, which is corrected for in APOKASC3 but not in our sample. This results in a measurement of slightly lower frequency (and therefore longer period) than the true $\nu_\mathrm{max}$. \textit{Right}: {Histograms of APOKASC3 periods (blue) and Kbonus periods (orange). We detect no periods in the 0--5 d bin due to our CNN detection limits, while 85\% of the APOKASC3 periods are in this bin (10\% are shorter than the 0.1 d edge of our wavelet transforms). Our CNN still detects fewer oscillators in the 5--10 d bin compared to APOKASC3, but we detect nearly all (plus a few extra) between 10 and 30 days. The Kbonus bins $>$ 35 d have 1 object each and are not significant enough to compare.}}
    \label{fig:oscillators}
\end{figure*}

The presence of oscillating giants in our sample provides the opportunity to compare the periods that our CNN recovered to the measured oscillation frequencies from asteroseismology. \Kepler\ oscillators have been the subject of a myriad of studies {\citep[e.g.,][]{Yu2018, Bedding2020}}, and oscillation frequencies are readily available. The APOGEE-\Kepler\ Asteroseismic Science Consortium (APOKASC) recently produced its third data release \citep{Pinsonneault2024}, providing high-quality asteroseismic parameters estimated from ten independent analyses using \Gaia\ luminosities and effective temperatures derived from high-resolution near-infrared spectroscopy. 

Our CNN found periods for 480 of the 15,808 giants analyzed by APOKASC3. In {the left panel of} Figure~\ref{fig:oscillators}, we plot our CNN period against the frequencies of maximum power ($\nu_\mathrm{max}$) found by APOKASC3. While our CNN was not designed to look for asteroseismic oscillations, we find remarkable agreement with the APOKASC frequencies, with bias of 2 days and mean scatter of 2.4 days. The bias is likely from the convective granulation background \citep{Bedding2020}, which APOKASC3 corrects for, but we do not. {We also examined the ranges of periods detected by APOKASC3 and our CNN, shown in the right panel of Figure~\ref{fig:oscillators}. The CNN performs best for periods between 10 and 30 days, detecting nearly all oscillators in APOKASC3, plus a few more.}

\subsection{Short Period Considerations}
\label{sec:short}
{As discussed by \citet{Claytor2022, Claytor2024, Ting2024}, and at the end of Section~\ref{sec:cnn}, one weakness of the CNN approach is the difficulty of measuring short periods. Our CNN often predicts incorrect periods for fast rotators, but most of them are successfully filtered out by their large uncertainties. This is evident in the middle panel of Figure~\ref{fig:predictions}, showing the results of evaluating the CNN on our simulated test set. In the region with True Period (x-axis) $<$ 3.5 days, many stars have Predicted Period (y-axis) up to 120 days, and the region has slightly higher number density than other “bad performance” regions. When we filter by the predicted uncertainty, most of these are removed, but as the first panel of Figure~\ref{fig:comparison} shows, some remain. \citet{McQuillan2014} found 2,888 stars with periods $<$ 3.5 days. Of these, we report ``good” periods for 83, but practically none of them match McQuillan. Of these 83 stars, \citet{Santos2019, Santos2021} reported periods for only 13, suggesting that the majority of the 83 McQuillan periods in this range may themselves be incorrect. Of the 13 in common, 12 periods match (within uncertainties) between Santos and McQuillan. For the one mismatch, Santos measured 7.5 days, and our measurement of 8.7 $\pm$ 3.7 days agrees with Santos. In summary, there will inevitably be some fast rotators that are mistakenly attributed long periods by the CNN method, but filtering by uncertainty is largely effective at removing them from the sample.}

\subsection{Comparison of Foreground and Background Periods}

\input{xcorr_table}

We obtained periods from both the foreground and background object for \Nbgfg\ pairs comprising 5,909 unique sources. The measurement of periods for both sources enables two tests: (1) to quantify the effectiveness of the PSF de-blending algorithm, and (2) to identify cases where a rotation period that was previously identified with the foreground object is actually associated with the background object. {Table~\ref{tab:xcorr} lists the pairs with key measurements, statistics, and flags used for the analysis in this section.}

\begin{figure*}
    \centering
    \includegraphics[width=0.8\linewidth]{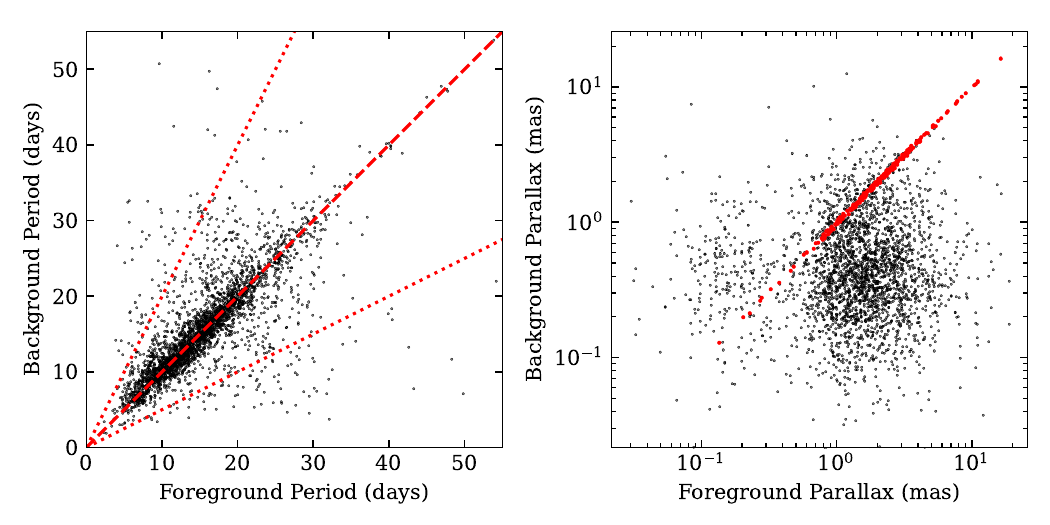}
    \caption{Rotation periods and parallaxes for \Nbgfg\ foreground-background pairs. \textit{Left:} Comparison of rotation periods of coincident foreground and background sources, with red lines denoting equality (dashed) and factor-of-two aliases (dotted). Most pairs have the same or nearly-the same period, suggesting that the light curves are not successfully de-blended. \textit{Right:} Parallax comparison for the same pairs. Pairs whose parallaxes are within 5\% of each other stand out and are plotted as red points. These \Nbin~pairs are likely to be binary systems.}
    \label{fig:pairs}
\end{figure*}

\pagebreak

\subsubsection{Assessing the quality of de-blending}

Figure~\ref{fig:pairs} shows the periods and parallaxes of the \Nbgfg~foreground-background pairs where a period was obtained for both members. 55\% of pairs had member periods within 10\% of one another (left panel), suggesting that the light curves are not successfully de-blended. Of the pairs, \Nbin~also have nearly the same parallax (right panel) and are likely to be binary systems. However, there are not enough binary candidates to account for the strong period correlation.

\begin{figure*}
    \centering
    \includegraphics[width=0.8\linewidth]{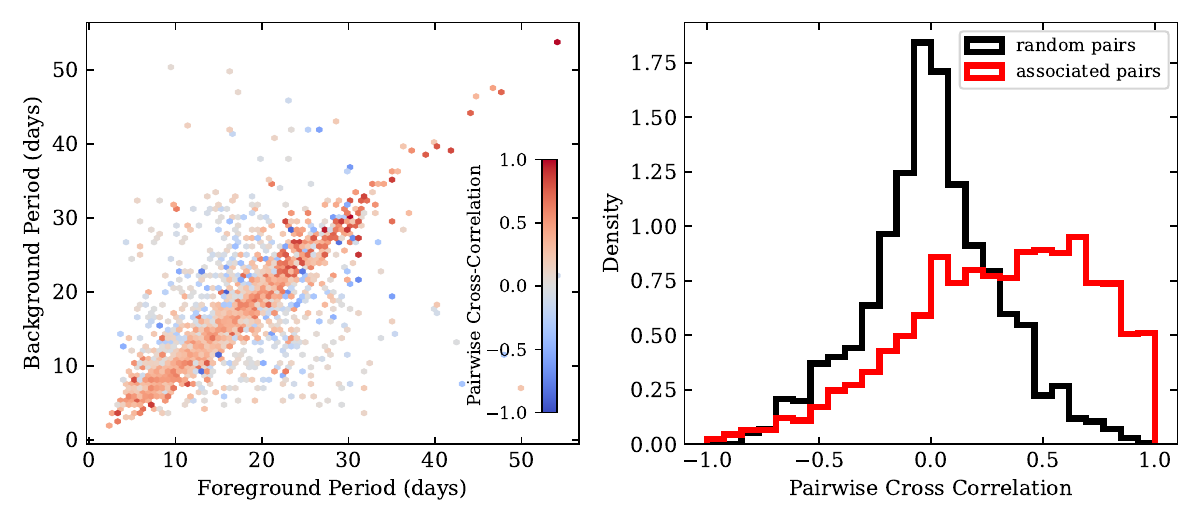}
    \caption{\textit{Left}: {The same as the left panel of Figure~\ref{fig:pairs}, but with bins colored by the light curve cross-correlation statistic. Pairs with the same period almost always have highly correlated light curves.} \textit{Right}: The distribution of cross-correlations between foreground-background light curves pairs (red) and 3,000 random pairs of unassociated light curves (black). As expected, the random pair cross-correlation distribution is concentrated around zero. However, the associated pairs' distribution is strongly skewed to positive correlation, with median cross-correlation of 0.35 (where 1 means perfect correlation and -1 means perfect anti-correlation). Thus, many background light curves are not successfully de-blended from their foreground neighbors.}
    \label{fig:xcorr}
\end{figure*}

To assess the quality of de-blending of light curve pairs for which we obtained periods, we computed the zero-lag cross-correlation or dot-product of each pair of light curves:
\begin{equation} \nonumber
    X = \frac{f(t) \cdot g(t)}{|f(t)| |g(t)|},
\end{equation}
where $f(t)$ and $g(t)$ are the paired light curves. By this definition, $X = 1$ denotes perfect correlation, and $X = -1$ denotes perfect anti-correlation. Note that this definition requires $f$ and $g$ to be defined at the same cadences. For our computation, we used only the candences where both $f$ and $g$ were defined. If a light curve pair had cadences that were mutually exclusive, the cross-correlation could not be computed, and a NaN value was assigned. This was the case for only 14 pairs. {The left panel of Figure~\ref{fig:xcorr} again shows the foreground and background source rotation periods, now colored by the light curve cross-correlation. Pairs with matching periods almost always have highly correlated light curves, suggesting that the pairs with period measurements are still highly blended. Interestingly, the cross-correlation statistic did not show significant trends with source separation, flux contrast, or Kbonus quality metrics such as PSFFRAC.}

Since two independent light curves should be neither correlated nor anticorrelated, we expect the distribution of cross-correlation statistics to be concentrated at zero. As a control, we also computed the cross-correlation between 3,000 random pairs of foreground and background light curves. To produce the pairs, we randomly selected without replacement 3,000 stars with defined KIC IDs as the ``foreground'' objects, and in the same fashion selected 3,000 stars without KIC IDs as the ``background'' objects. Because the stars in a pair are randomly distributed across the \Kepler\ field, their light curves should have no causal relationship, and the distribution of cross-correlation values represents a null distribution. 

The cross-correlation distributions are shown in {the right panel of} Figure~\ref{fig:xcorr}. The random pairs (black) are concentrated about zero as expected, but the associated pairs (red) are much more often positively correlated, with a median cross-correlation of 0.35. Comparing each cross-correlation value $X$ to the null distribution, we computed single-tail $p$-values indicating the probability that a given pair of light curves was correlated (or anti-correlated). The summary statistics are reported in Table~\ref{tab:stats}. Of the light curve pairs for which we detected periods for both sources, 53\% are positively correlated at the $p < 0.15$ (roughly 1$\sigma$) level, and 10\% are negatively correlated at the same level. This suggests that, at least among light curves with strong periodic variability, the majority of foreground-background pairs are not successfully de-blended. \citet{Martinez-Palomera2023} similarly found that de-blending targets was more difficult in cases with strong variability, especially when the variable source is much brighter. Since our sample is composed entirely of periodic signals, it is likely that strongly correlated pairs are overrepresented compared to the full Kbonus data set.

\input{stats_table}

\begin{figure}
    \centering
    \includegraphics[width=\linewidth]{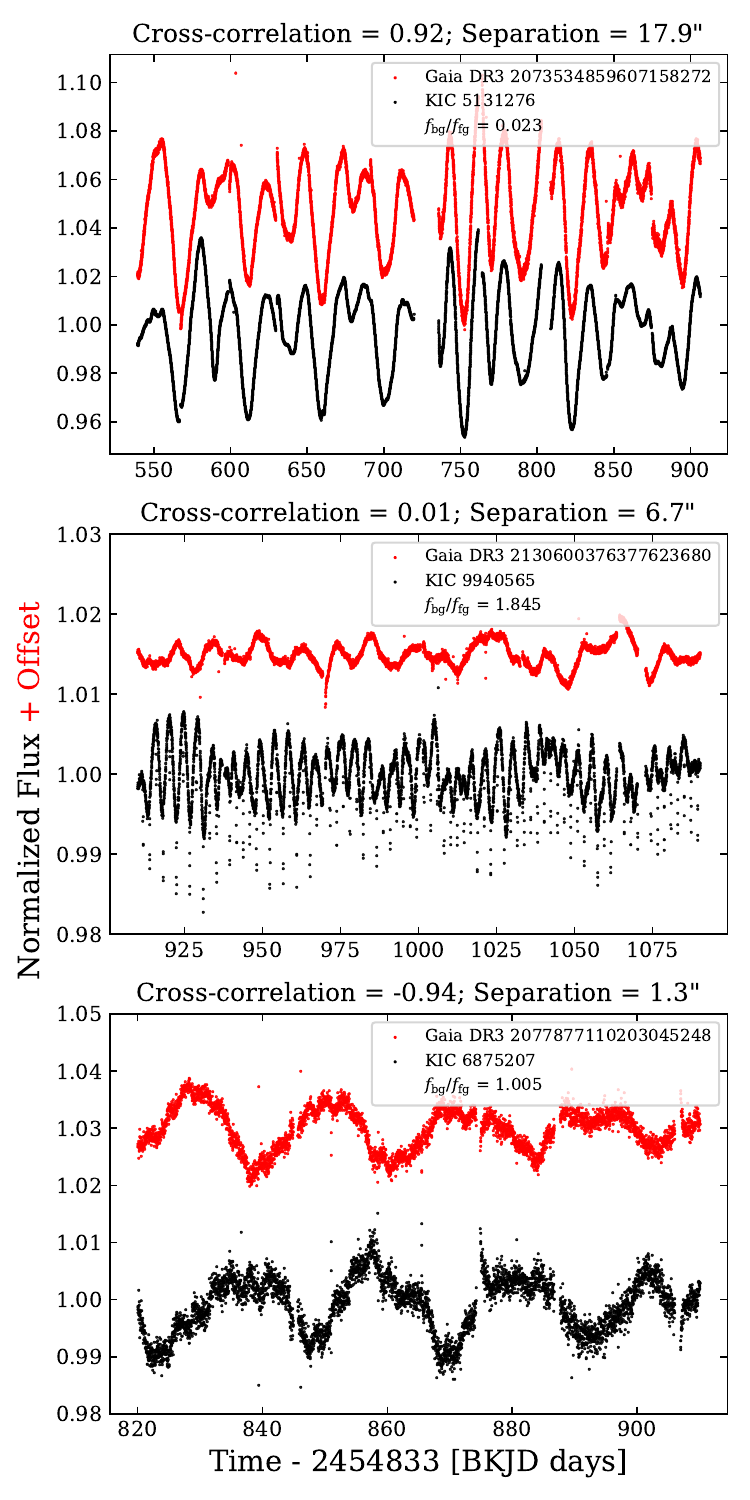}
    \caption{Three examples of foreground-background pair light curves with different levels of cross-correlation. The foreground source light curve is shown in black, and the background light curve in red. \textit{Top}: highly correlated light curves that appear not to be successfully de-blended. The two light curves are nearly identical despite being separated by nearly 18\arcsec, owing to the foreground star being over 40 times brighter than the background star. \textit{Middle}: a light curve pair with nearly zero correlation which appears to be successfully de-blended. \textit{Bottom}: a highly anti-correlated light curve pair.}
    \label{fig:lcs_xcorr}
\end{figure}

Figure~\ref{fig:lcs_xcorr} shows examples of light curve pairs with strong correlation ($X = 0.92$), weak correlation ($X = 0.01$), and strong anti-correlation ($X = -0.94$). The light curves in the strongly correlated pair appear to be nearly identical in shape, suggesting that the de-blending algorithm failed for this source. In the weakly correlated case, the light curves appear to be completely independent despite being near each other on the detector, showing two different rotation periods, and the foreground object exhibiting what are likely to be binary eclipses. In this case, the light curves were successfully de-blended, allowing us to obtain a high-fidelity rotation period of the background source. 

The strongly anti-correlated light curves exhibit nearly the same rotation period, but are $180^\circ$ out of phase with one another. Several scenarios could explain this behavior. One possibility is that the light curves were successfully de-blended and truly have similar period and opposite phase. This could be a coincidence, or it could be that the stars are physically associated, tidally synchronized, and have active regions opposite one another due to magnetic interactions. However, the stars in this example are separated by over 50 pc according to their \Gaia\ parallaxes, ruling out the latter scenario. A more likely possibility is that the PSF de-blending algorithm associated too much flux with one target, enhancing its rotation signature while creating an opposite signature in the other light curve. \texttt{PSFMachine} uses least-squares methods to solve the linear model $\mathbf{f}= \mathrm{X} \mathbf{w}$, where $\mathbf{f}$ is the flux, X is a design matrix, and $\mathbf{w}$ is the vector of best-fit coefficients \citep{Hedges2021}. There are no mathematical constraints besides Gaussian priors, allowing for anticorrelated solutions that effectively minimize the model in blended sources. Anti-correlated light curves ($X \lesssim 0.85$) are slightly over-represented in the associated pairs when compared to the null distribution (Figure~\ref{fig:xcorr}), suggesting that some over-correction is likely present. 

\subsubsection{Source confusion}

\begin{figure}
    \centering
    \includegraphics[width=\linewidth]{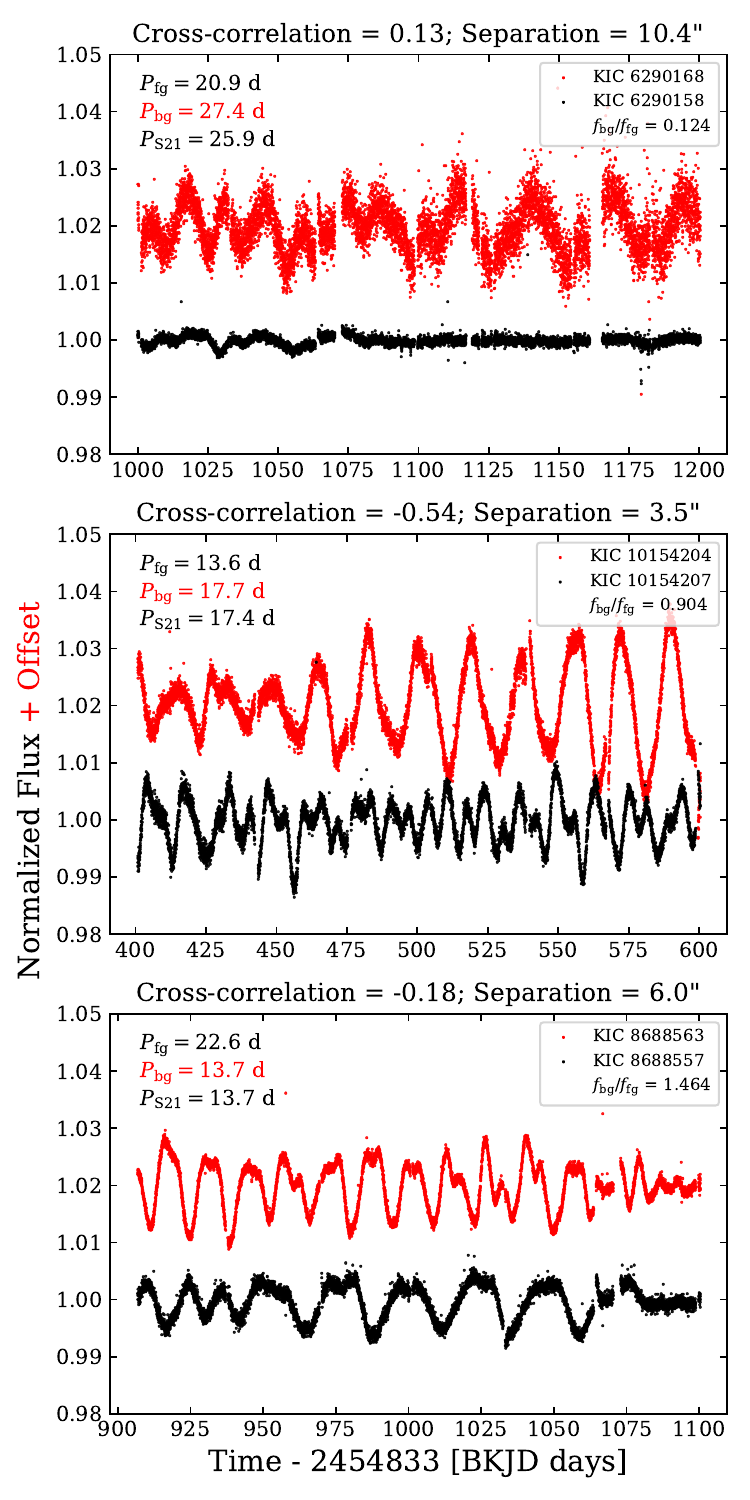}
    \caption{Three examples of de-blended foreground-background pair light curves where the rotation period was previously attributed to the foreground source, but is now identified with the background source. The foreground source light curve is shown in black, and the background light curve in red. The text in the upper left of each panel shows the foreground source period, the background source period, and the period measured by \citet{Santos2019, Santos2021}.}
    \label{fig:lcs_src}
\end{figure}

Finally, we investigated cases where a period that was previously associated with a foreground source was now found to be associated with a background source. We required these pairs to satisfy the following criteria:
\begin{itemize}
    \item both sources in a pair must have a period,
    \item the two periods must differ by at least 10\%,
    \item the two periods must not be within 10\% of a factor-of-two alias,
    \item the two sources must not be binary companions, i.e., their parallaxes must differ by at least 5\%, and
    \item the background period must be within 10\% of the foreground source's period from \citet{Santos2019, Santos2021}.
\end{itemize}

Applying these criteria, we identified {220} candidate pairs where the period was previously confused with the foreground source. Figure~\ref{fig:lcs_src} shows three examples of such pairs. The light curves are visibly uncorrelated and clearly display different rotation signatures, even when the sources are separated by less than a pixel width, highlighting the ability of the \texttt{psfmachine} to reduce source confusion in crowded fields or large-pixel detectors. 

\section{Conclusion}
We used the deep learning framework of \citet{Claytor2022, Claytor2024} to obtain \Nperiods\ stellar rotation periods from the KBonus-Background light curves \citep{Martinez-Palomera2023}. \Nallnew\ periods were newly measured from stars which lacked previous measurements. We reproduced the \Kepler\ period distribution seen by \citet{McQuillan2014, Santos2019, Santos2021, Reinhold2023}, including the short-period edge, the intermediate-period gap, the and the slowly-rotating M dwarf ``tail". Comparing our periods of \Kepler\ prime sources to those measured in the literature, we found excellent agreement with autocorrelation- and wavelet-based methods \citep{McQuillan2014, Santos2019}, and good agreement {with power spectrum gradient techniques \citep{Reinhold2023}} except for a cluster of half-period alises. We also obtained oscillation periods for \Npulsators\ asteroseismic giant stars with 0.30 $\mu$Hz $< \nu_\mathrm{max} <$ 6.1 $\mu$Hz and 0.33 $< \logg <$ 1.7, which agree very well with the most carefully measured oscillation frequencies from the literature.

The recovery of \Kepler\ rotation periods provides the first real-world validation test of the deep learning framework. The framework was developed on simulated light curves from TESS, for which no large period catalog existed at the time. The accuracy with which we recovered literature periods establishes CNNs as a viable class of methods for determining modulation periods en masse. Simultaneously, we have extended the period sample from \Kepler\, providing rotation periods for previously unmeasured sources, as well as oscillation frequencies our CNN was not designed to detect.

Comparing the periods and light curves of foreground and background sources provided a way to quantify the effectiveness of the \texttt{psfmachine} algorithm \citep{psfmachine} at reducing source confusion. Importantly, we found that
\begin{itemize}
    \item foreground-background pairs often had the same period (55\% of pairs had periods agreeing to within 10\%), even when removing binary systems from the comparison, suggesting that contamination is still present in many sources;
    \item the single-lag cross correlation (or normalized dot product) of two light curves effectively identifies source pairs whose light curves are likely still blended;
    \item as many as 63\% of periodic KBonus-Background light curves remain blended with their foreground counterparts; and
    \item when successfully de-blended, the background light curves enable independent period detection and the reduction of source confusion.
\end{itemize}

The Kbonus background light curves give new life to the thoroughly-studied \Kepler\ data. Access to background source light curves can provide an unbiased sample for comparison with \Kepler's primary targets, which had a complex, multi-stage selection function. However, at this stage, periodic light curves are not sufficiently de-blended from their foreground counterparts to enable science at scale with such an unbiased sample. With a more de-blended background sample, we propose the following tests for future work:

\begin{itemize}
    \item Testing gyrochronology with wide-binary pairs: do stars in wide binaries have the same gyrochronological age?
    \item Testing the effects of binarity on rotational evolution: how does the protostellar disk structure in a binary system affect the early evolution of rotation?
    \item Testing the distribution of background rotation periods vs foreground rotation periods: does the foreground sample show biases, and how do these biases affect, e.g., the inferred age distributions, expectations for planet occurrence, etc.?
    \item Testing truncated rotational braking: do background stars undergo truncated braking at the same Rossby number as foreground stars?
    \item Testing the period gap: is the rotation period gap at the same periods for the background stars as for the foreground stars?
\end{itemize}

More generally, combining precise position information with advanced linear methods can help de-blend sources and dramatically increase the number of targets that are accessible to current missions (TESS) and upcoming missions (PLATO, Roman, Rubin). This is especially important where crowding and source confusion pose significant threats, such as the large pixel scale of TESS or the dense crowding of Roman's Galactic Bulge Time Domain Survey. As de-blending methods mature, we urge that care be taken when applying them and interpreting their results since residual contamination may introduce complex biases that are difficult to account for.

\vfill\null

\section{Acknowledgments} 
The authors thank the anonymous reviewer for thorough and thoughtful feedback that improved the quality of this manuscript.

We thank Jorge Mart{\'\i}nez-Palomera and Christina Hedges for helpful insights into the Kbonus data, Jinmi Yoon and Deborah Kenny for help mirroring the Kbonus data set, Karolina Garcia and Adam Ginsburg for help with \texttt{AstroPy} troubleshooting, Gagandeep Anand, Erica Bufanda, Lyra Cao, Ryan Dungee, Meir Schochet, Michele Silverstein, and the research group of Rana Ezzedine and Jamie Tayar at the University of Florida Department of Astronomy for fruitful conversations that improved our analysis.

We acknowledge University of Florida Research Computing for providing computational resources and support that have contributed to the research results reported in this publication. URL: \url{https://www.rc.ufl.edu}.

The authors acknowledge support from the National Aeronautics and Space Administration (grant No. 80NSSC24K0081).

This paper includes data collected by the Kepler mission and obtained from the MAST data archive at the Space Telescope Science Institute (STScI). Funding for the Kepler mission is provided by the NASA Science Mission Directorate. STScI is operated by the Association of Universities for Research in Astronomy, Inc., under NASA contract NAS 5–26555.

This work has made use of data from the European Space Agency (ESA) mission \Gaia\ (\url{https://www.cosmos.esa.int/gaia}), processed by the \Gaia\ Data Processing and Analysis Consortium (DPAC, \url{https://www.cosmos.esa.int/web/gaia/dpac/consortium}). Funding for the DPAC has been provided by national institutions, in particular the institutions participating in the \Gaia\ Multilateral Agreement.

\software{
    \texttt{butterpy} \citep{butterpy1.0, Claytor2022}, 
    \texttt{Lightkurve} \citep{Lightkurve2018}, 
    \texttt{AstroPy} \citep{Astropy2013, Astropy2018, Astropy2022},
    \texttt{Astroquery} \citep{Ginsburg2019},
    \texttt{iPython} \citep{iPython2007}, 
    \texttt{Matplotlib} \citep{Matplotlib2007}, 
    \texttt{NumPy} \citep{Numpy2020}, 
    \texttt{Pandas} \citep{Pandas2010}, 
    \texttt{PyTorch} \citep{Pytorch2019}, 
    \texttt{SciPy} \citep{Scipy2020}. 
    }

\pagebreak 

\bibliography{references}{}
\bibliographystyle{aasjournal}

\end{document}

%% file: periods_table.tex
\begin{deluxetable}{ll}
    \tabletypesize{\footnotesize}
    \tablecaption{Periods of Kbonus Sources}
    \tablehead{\colhead{Column} & \colhead{Description}}
    \startdata
        gaia\_designation   & Gaia DR3 designation \\
        ra                  & Gaia DR3 right ascension (deg) \\
        dec                 & Gaia DR3 declination (deg) \\
        phot\_g\_mean\_mag  & Gaia DR3 $G$ magnitude \\
        phot\_bp\_mean\_mag & Gaia DR3 $BP$ magnitude \\
        phot\_rp\_mean\_mag & Gaia DR3 $RP$ magnitude \\
        parallax            & Gaia DR3 parallax (mas) \\
        ruwe                & Gaia DR3 renormalized unit weight error \\
        kic                 & Kepler Input Catalog designation \\
        kep\_mag            & Kepler magnitude \\
        nquarters           & Number of quarters used from light curve \\
        period              & CNN-inferred period (days) \\
        period\_err         & period uncertainty (days) \\
        pulsator\_flag      & flag denoting likely pulsating giants \\
        psffrac             & * fraction of PSF captured in the target \\
                            & pixel file \\
        pertrati            & * ratio of mean model flux to perturbed \\
                            & model flux \\
        pertstd             & * ratio of mean model flux standard \\
                            & deviation to that of the perturbed model \\
    \enddata
    \tablecomments{Only the columns of this table are shown here to demonstrate its form and content. A machine-readable version of the full table is available online. Column descriptors marked by asterisks (*) are reproduced from \citet{Martinez-Palomera2023}.}
    \label{tab:periods}
\end{deluxetable}

%% file: xcorr_table.tex
\begin{deluxetable}{ll}
    \tabletypesize{\footnotesize}
    \tablecaption{Foreground-Background Pairs}
    \tablehead{\colhead{Column} & \colhead{Description}}
    \startdata
        gaiadr3\_bg          & Background source Gaia DR3 ID \\
        kic\_bg              & Background source KIC ID \\
        psf\_mean\_flux\_bg  & Background source mean flux (e$^-$/s) \\
        parallax\_bg         & Background source Gaia DR3 parallax (mas) \\
        period\_bg           & Background source period (days)\\
        period\_err\_bg      & Background source period uncertainty (days) \\
        gaiadr3\_fg          & Foreground source Gaia DR3 ID \\
        kic\_fg              & Foreground source KIC ID \\
        psf\_mean\_flux\_fg  & Foreground source mean flux (e$^-$/s) \\
        parallax\_fg         & Foreground source Gaia DR3 parallax (mas) \\
        period\_fg           & Foreground source period (days) \\
        period\_err\_fg      & Foreground source period uncertainty (days) \\
        xcorr                & Light curve cross-correlation \\
        sep                  & Source separation (\arcsec) \\
        flag\_match          & Flag for matching period candidates \\
        flag\_alias          & Flag for period alias candidates \\
        flag\_binary         & Flag for binary pair candidates \\
        flag\_source         & Flag for source confusion candidates \\
    \enddata
    \tablecomments{Only the columns of this table are shown here to demonstrate its form and content. A machine-readable version of the full table is available online.}
    \label{tab:xcorr}
\end{deluxetable}

%% file: stats_table.tex

\begin{deluxetable}{ccc}
    \tablecaption{Correlated Light Curve Pair Statistics}
    \tablehead{\colhead{\hspace{0.1cm}$p$-value range}\hspace{0.4cm} & \colhead{\hspace{0.4cm}$f_{\mathrm{pos}}$ (\%)}\hspace{0.4cm} & \colhead{\hspace{0.4cm}$f_{\mathrm{neg}}$ (\%)}\hspace{0.1cm}}
    \startdata
        $0.15 > p \geq 0.02$ & 30 & 7 \\
        $0.02 > p \geq 0.01$ & 7 & 1 \\
        $0.01 > p \geq 0.001$ & 9  & 1 \\
        $0.001 > p \geq 0.0001$ & 5  & 0 \\
        $p < 0.0001$ & 2  & 1 
    \enddata
    \tablecomments{The percentages of light curve pairs that are correlated ($f_\mathrm{pos}$) or anti-correlated ($f_\mathrm{neg}$) at various probability levels. Summed together, as many as 63\% of light curve pairs with rotation periods are likely to remain blended at least at the $p < 0.15$ (roughly 1$\sigma$) level.}
    \label{tab:stats}
\end{deluxetable}